\documentclass[aps,prl,twocolumn,amsmath,amssymb]{revtex4}
\usepackage{dcolumn}
\usepackage{bm}
\usepackage{graphicx}
\begin{document}
\title{Comment on ``Formation of magnetic glass in calcium doped YBaCo$_2$O$_{5.5}$ cobaltites" [arXiv:1106.6253].}
\author{P. Chaddah and A. Banerjee}
\affiliation{UGC-DAE Consortium for Scientific Research, University Campus, Khandwa Road, Indore-452001, Madhya Pradesh, India.}

\begin{abstract}
A recent paper from Raveau's group (arXiv:1106.6253) has shown the existence of a glass-like arrested state in two calcium-doped cobaltites, which consists of coexisting ferromagnetic and antiferromagnetic phase fractions, resulting from arrest of kinetics at low temperature,. They use the ``cooling and heating in unequal magnetic field (CHUF)" measurement protocol to show devitrification of the arrested state. They use the same section title, protocol name, and acronym as in the paper (arXiv:0805.1514) where the protocol was presented, justified, and exploited. They nowhere refer to that paper, or to its journal counterpart. We put the development of this CHUF protocol in perspective, bringing out that it was conceived for broadened first order transitions where the kinetics of the transformation gets hindered. We also point out its applicability in view of our various subsequent works (arXiv:0911.4552, arXiv:1003.4400 and their journal counterparts) further exploiting this protocol. The paper being commented on is also silent about these latter papers (or their journal counterparts) even though the use of this protocol is highlighted in these  in the abstract itself.  
\end{abstract}

\maketitle
The kinetics of a first order transition is dictated by the time required for the latent heat to be extracted. This is dictated by intrinsic processes like diffusion of molecules to attain the density, and the order, of the solid when the transition corresponds to solidification of a liquid. Assuming that the former is slower and thus dictates the time scale, rapid quenching (or splat cooling) was discovered as a technique to make metallic glasses. This time scale would grow if the change in density is to occur at a lower temperature. Many years back, Manekar et al. \cite{manekar} proposed that magnetic first order transitions would similarly require finite time. The time scale would again increase as temperature is lowered, if a structural change (with change in density) occurred simultaneously. They explained a lot of their perplexing data by proposing that the magnetic transition proceeded over a temperature (or magnetic field) range, due to disorder-induced broadening, and a part of the transition was completed while the remaining part was kinetically hindered over experimental time scales. They referred to this as arrest of kinetics, and postulated that the low-temperature state was a glass-like arrested metastable high-temperature phase coexisting with an equilibrium low-temperature phase. They stated in their abstract that this could also explain some of the observations of Tokura's group in half-doped manganites. This work continued with the additional measurement of relaxation, the observation of kinetic arrest was extended to some more magnetic materials, and the term `magnetic glass' was introduced in this context \cite{roy}. The many publications during this period by various authors from that group, and their collaborators, can be obtained from reference \cite{roy} and from references therein. The paper being commented on only refers to \cite{roy}, and not to any earlier work or even to `references therein', thus denying readers information on what the term `magnetic glass' really connotes.

The next development in this saga was arXiv:0601095 which addressed the question of whether there was any correlation between how disorder-induced broadening affected the transition temperature T$_C$ and the kinetic arrest temperature T$_K$, restricting itself to the two extreme monotonic options termed as `correlation' or `anti-correlation' of the two broadened bands \cite{chaddah}. This led seamlessly to the realization \cite{ban1} that (i) the phase fractions of the two phases coexisting at the lowest temperature could be tuned by varying the cooling field, and that (ii) cooling and heating in different fields (H$_C$ and H$_W$ respectively) would result in a reentrant transition for only one sign of (H$_C$-H$_W$). The observation of `kinetic arrest' was being reported in newer materials, and our group presented, justified, and outlined the potential of such a measurement protocol, giving it a name and an acronym \cite{arxiv}. The measurement protocol CHUF was then exploited in some of our further studies \cite{pallavi}. This presents our perspective view and counters the impression, created by the authors of the paper being commented on, of their originality regarding the CHUF protocol. This is an essential supplement to the paper being commented on since neither the initial concept i.e. Ref. [1] nor the development of the CHUF protocol Refs. [3-6] have been cited in that paper. 

The `accepted papers' site of Physical Review B is showing that an Erratum to the journal counterpart of this paper was accepted for publication on July 25, and the contents of this Erratum are accepted/decided by the journal editors. However, the authors have not responded to our Comment on this arXiv paper. It is well recognized that with the present policy of Physical Review B, ``pointing out this type of omission in an Erratum means that it is less likely to be found and cited" \cite{prb}. The arXiv site is less friendly to any attempt at usurping credit because such an omission is to be corrected by modifying the uploaded version and all versions are available with the same access number. This is to be understood in conjunction with Martin Blume's editorial \cite{blume} where authors are urged to make ``a prompt correction" to omission of references and it is noted that this failure ``can cross the line to plagiarism". The absence of a reply to our Comment, and the lack of uploading of v2 of the paper being commented on, is a serious matter of violation of ethics.

\end{document}